\documentclass[aps,prd,twocolumn,showpacs]{revtex4-1}

\usepackage{wallpaper}
\usepackage[colorlinks,linkcolor=blue,anchorcolor=blue,citecolor=red]{hyperref}
\usepackage{amsmath}
\usepackage{multirow}

\begin{document}
\hyphenpenalty=6000
\tolerance=1000

\hyphenation{Eq}

\title{Unified neutron star EOSs and neutron star structures in RMF models}

\author{Cheng-Jun~Xia$^{1,2,3}$}
\email{cjxia@yzu.edu.cn}
\author{Toshiki Maruyama$^{3}$}
\email{maruyama.toshiki@jaea.go.jp}
\author{Ang Li$^{4}$}
\email{liang@xmu.edu.cn}
\author{Bao Yuan Sun$^{5, 6}$}
\email{sunby@lzu.edu.cn}
\author{Wen-Hui Long$^{5, 6}$}
\email{longwh@lzu.edu.cn}
\author{Ying-Xun Zhang$^{7, 8}$}
\email{zhyx@ciae.ac.cn}

\affiliation{$^{1}${Center for Gravitation and Cosmology, College of Physical Science and Technology, Yangzhou University, Yangzhou 225009, China}
\\$^{2}${School of Information Science and Engineering, NingboTech University, Ningbo 315100, China}
\\$^{3}${Advanced Science Research Center, Japan Atomic Energy Agency, Shirakata 2-4, Tokai, Ibaraki 319-1195, Japan}
\\$^{4}${Department of Astronomy, Xiamen University, Xiamen 361005, China}
\\$^{5}${School of Nuclear Science and Technology, Lanzhou University, Lanzhou 730000, China}
\\$^{6}${Frontiers Science Center for Rare Isotopes, Lanzhou University, Lanzhou 730000, China}
\\$^{7}${China Institute of Atomic Energy, Beijing 102413, People's Republic of China}
\\$^{8}${Guangxi Key Laboratory Breeding Base of Nuclear Physics and Technology, Guilin 541004, China}
}

\date{\today}

\begin{abstract}
In the framework of Thomas-Fermi approximation, we study systematically the EOSs and microscopic structures of neutron star matter in a vast density range with $n_\mathrm{b}\approx 10^{-10}$-2 $\mathrm{fm}^{-3}$, where various covariant density functionals are adopted, i.e., those with nonlinear self couplings (NL3, PK1, TM1, GM1, MTVTC) and density-dependent couplings (DD-LZ1, DDME-X, PKDD, DD-ME2, DD2, TW99). It is found that the EOSs generally coincide with each other at $n_\mathrm{b}\lesssim 10^{-4}$ fm${}^{-3}$ and 0.1 fm${}^{-3}\lesssim n_\mathrm{b} \lesssim 0.3$ fm${}^{-3}$, while in other density regions they are sensitive to the effective interactions between nucleons. By adopting functionals with larger slope of symmetry energy $L$, the curvature parameter $K_\mathrm{sym}$ and neutron drip density generally increase, while the droplet size, proton number of nucleus, core-crust transition density, and onset density of non-spherical nuclei decrease. All functionals predict neutron stars with maximum masses exceeding the two-solar-mass limit, while those of DD2, DD-LZ1, DD-ME2, and DDME-X predict optimum neutron star radii according to the observational constraints. Nevertheless, the corresponding skewness coefficients $J$ are much lager than expected, while only the functionals MTVTC and TW99 meet the start-of-art constraints on $J$. More accurate measurements on the radius of PSR J0740+6620 and the maximum mass of neutron stars are thus essential to identify the functional that satisfies all constraints from nuclear physics and astrophysical observations. Approximate linear correlations between neutron stars' radii at $M=1.4 M_{\odot}$ and $2 M_{\odot}$, the slope $L$ and curvature parameter $K_\mathrm{sym}$ of symmetry energy are observed as well, which is mainly attributed to the curvature-slope correlations in the functionals adopted here. The results presented here are applicable for the investigations on the structures and evolutions of compact stars in a unified manner.
\\

\end{abstract}

\maketitle

\section{\label{sec:intro}Introduction}
Due to the challenges reside in simulating dense matter with lattice QCD, the state and composition of stellar matter inside compact stars is still unclear and exhibits large ambiguities. In particular, the uncertainties in the equation of state (EOS) and the corresponding microscopic structures are still sizable~\cite{Dutra2012_PRC85_035201, Dutra2014_PRC90-055203, Xia2020_PRD102-023031, LI2020, Hebeler2021_PR890-1}, which was shown to play important roles in the properties and evolutions of compact stars~\cite{Pons1999_ApJ513-780, Horowitz2004_PRC69-045804, Lattimer2012_ARNPS62-485, Janka2012_ARNPS62-407, Bauswein2012_PRD86-063001, Rueda2014_PRC89-035804, Watanabe2017_PRL119-062701, Sotani2019_MNRAS489-3022, Koeppel2019_ApJ872-L16, Baiotti2019_PPNP109-103714, Schuetrumpf2020_PRC101-055804, Bauswein2020_PRL125-141103, Gittins2020_PRD101-103025, Preau2021_MNRAS505-939}. The properties of nuclear matter around the saturation density ($n_0\approx 0.16\ \mathrm{fm}^{-3}$), on the contrary, are well constrained with various terrestrial experiments, astrophysical observations, and nuclear theories, where the binding energy is $B\approx -16$ MeV, the incompressibility $K = 240 \pm 20$ MeV~\cite{Shlomo2006_EPJA30-23}, the symmetry energy $S = 31.7 \pm 3.2$ MeV and its slope $L = 58.7 \pm 28.1$ MeV~\cite{Li2013_PLB727-276, Oertel2017_RMP89-015007}. Note that those quantities can be further constrained by considering the up-to-date astrophysical observations, heavy ion collision data, measurements of the neutron skin thickness for $^{208}$Pb in PREX-II~\cite{PREX2021_PRL126-172502}, as well as predictions of chiral effective field theory, e.g., those in Refs.~\cite{Zhang2020_PRC101-034303, Essick2021_PRL127-192701}. Meanwhile, at vanishing densities, the interaction among nucleons is negligible so that nuclear matter exhibits gas phase with well understood properties~\cite{Yang2019_PRC100-054314, Yang2021_PRC103-014304}.

At subsaturation densities, due to the liquid-gas phase transition of nuclear matter, a mixed phase with various nonuniform structures is expected, i.e., nuclear pasta~\cite{Baym1971_ApJ170-299, Negele1973_NPA207-298, Ravenhall1983_PRL50-2066, Hashimoto1984_PTP71-320, Williams1985_NPA435-844}, which is typically found in the crusts of neutron stars and the core region of supernovae at the stage of gravitational collapse. By employing spherical and cylindrical approximations for the Wigner-Seitz (WS) cell~\cite{Pethick1998_PLB427-7, Oyamatsu1993_NPA561-431, Maruyama2005_PRC72-015802, Togashi2017_NPA961-78, Shen2011_ApJ197-20}, five types of geometrical structures for nuclear pasta were obtained aside from the uniform phase, i.e, droplets, rods, slabs, tubes, and bubbles, while further investigations have revealed more complicated structures~\cite{Magierski2002_PRC65-045804, Watanabe2003_PRC68-035806, Newton2009_PRC79-055801, Nakazato2009_PRL103-132501, Okamoto2012_PLB713-284, Schuetrumpf2013_PRC87-055805, Schneider2014_PRC90-055805, Schuetrumpf2015_PRC91-025801,  Fattoyev2017_PRC95-055804, Schuetrumpf2019_PRC100-045806, Sagert2016_PRC93-055801, Berry2016_PRC94-055801, Kashiwaba2020_PRC101-045804}. Nevertheless, the investigation on the EOSs and microscopic structures of nuclear pasta is still far from complete due to the uncertainties in the nuclear energy density functional~\cite{Douchin2001_AA380-151, Sharma2015_AA584-A103, Fortin2016_PRC94-035804, Pearson2018_MNRAS481-2994, Vinas2021_Symmetry13-1613, DinhThi2021_AA654-A114, Newton2022_EPJA58-69}, while a unified treatment is preferred so that the uncertainties do not get larger~\cite{Fortin2016_PRC94-035804, DinhThi2021_AA654-A114}.

For stellar matter at larger densities, as we are entering the multimessenger era, constraining the EOS with pulsar observations has reached unprecedent accuracy. For example, the observation of two-solar-mass pulsars~\cite{Demorest2010_Nature467-1081, Antoniadis2013_Science340-1233232, Fonseca2016_ApJ832-167, Cromartie2020_NA4-72, Fonseca2021_ApJ915-L12} have excluded various soft EOSs for dense stellar matter. The multi-messenger observations of the binary neutron star merger event {GRB} 170817A-{GW}170817-{AT} 2017gfo have constrained the tidal deformability of $1.4 M_{\odot}$ neutron star with $70\leq \Lambda_{1.4}\leq 580$ and the radii $R=11.9\pm1.4$ km~\cite{LVC2018_PRL121-161101}, indicating a soft EOS at small densities. Additionally, based on pulse-profile modeling with {NICER} and {XMM}-Newton data, the simultaneous measurements of the masses and radii for PSR J0030+0451 and PSR J0740+6620~\cite{Riley2019_ApJ887-L21, Riley2021_ApJ918-L27, Miller2019_ApJ887-L24, Miller2021_ApJ918-L28} suggest that their radii are similar ($\sim$12.4 km) despite the large differences in masses. In such cases, the likelihood of a strong first-order phase transition inside two-solar-mass pulsars may be reduced~\cite{Pang2021_ApJ922-14}.

The purpose of our current study is twofold. First, we examine the structures of neutron stars without introducing any new degrees of freedom that leads to first-order phase transitions. Since the radius and crust thickness of a neutron star are sensitive to the EOSs~\cite{Fortin2016_PRC94-035804}, a unified description for neutron star matter is thus necessary~\cite{Fortin2016_PRC94-035804, DinhThi2021_AA654-A114}. This leads to the second purpose of our study, where we have obtained 11 EOSs and the corresponding microscopic structures of neutron star matter in a unified manner adopting the numerical recipe proposed in Ref.~\cite{Xia2022_PRC105-045803}. In particular, as was done in Refs.~\cite{Maruyama2005_PRC72-015802, Avancini2008_PRC78-015802, Avancini2009_PRC79-035804, Gupta2013_PRC87-028801}, the properties of nuclear matter are fixed with relativistic mean field (RMF) models~\cite{Meng2016_RDFNS}, which was very successful in describing finite nuclei~\cite{Reinhard1989_RPP52-439, Ring1996_PPNP37_193-263,
Meng2006_PPNP57-470, Paar2007_RPP70-R02, Meng2015_JPG42-093101, Meng2016_RDFNS, Chen2021_SCPMA64-282011, Typel1999_NPA656-331, Vretenar1998_PRC57-R1060, Lu2011_PRC84-014328} and nuclear matter~\cite{Glendenning2000, Ban2004_PRC69-045805, Weber2007_PPNP59-94, Long2012_PRC85-025806, Sun2012_PRC86-014305, Wang2014_PRC90-055801, Fedoseew2015_PRC91-034307, Gao2017_ApJ849-19}. Two types of RMF Lagrangian are considered, i.e., those with nonlinear self couplings (NL3~\cite{Lalazissis1997_PRC55-540}, PK1~\cite{Long2004_PRC69-034319}, TM1~\cite{Sugahara1994_NPA579-557}, GM1~\cite{Glendenning1991_PRL67-2414}, MTVTC~\cite{Maruyama2005_PRC72-015802}) and density-dependent couplings (DD-LZ1~\cite{Wei2020_CPC44-074107}, DDME-X~\cite{Taninah2020_PLB800-135065}, PKDD~\cite{Long2004_PRC69-034319}, DD-ME2~\cite{Lalazissis2005_PRC71-024312}, DD2~\cite{Typel2010_PRC81-015803}, TW99~\cite{Typel1999_NPA656-331}).

The paper is organized as follows. In Sec.~\ref{sec:the} we present the theoretical framework for the covariant density functionals adopted here and fixing the microscopic structures of neutron star matter. The obtained EOSs and microscopic structures of neutron star matter are presented in Sec.~\ref{sec:results}, while the corresponding neutron star structures and the possible correlations with the symmetry energy coefficients are investigated. We draw our conclusion in Sec.~\ref{sec:con}

\section{\label{sec:the} Theoretical framework}

\subsection{\label{sec:the_RMF} RMF models}
The Lagrangian density of RMF models for the neutron star matter considered here reads
\begin{eqnarray}
\mathcal{L}
 &=& \sum_{i=n,p} \bar{\psi}_i
       \left[  i \gamma^\mu \partial_\mu - \gamma^0 \left(g_\omega\omega + g_\rho\rho\tau_i + A q_i\right)- m_i^* \right] \psi_i
\nonumber \\
 &&\mbox{} + \sum_{l=e,\mu} \bar{\psi}_l \left[ i \gamma^\mu \partial_\mu - m_l + e \gamma^0 A \right]\psi_l - \frac{1}{4} A_{\mu\nu}A^{\mu\nu}
\nonumber \\
 &&\mbox{} + \frac{1}{2}\partial_\mu \sigma \partial^\mu \sigma  - \frac{1}{2}m_\sigma^2 \sigma^2
           - \frac{1}{4} \omega_{\mu\nu}\omega^{\mu\nu} + \frac{1}{2}m_\omega^2 \omega^2
\nonumber \\
 &&\mbox{} - \frac{1}{4} \rho_{\mu\nu}\rho^{\mu\nu} + \frac{1}{2}m_\rho^2 \rho^2 + U(\sigma, \omega),
\label{eq:Lagrange}
\end{eqnarray}
where $\tau_n=-\tau_p=1$ is the 3rd component of isospin, $q_i=e (1-\tau_i)/2$ the charge, and $m_{n,p}^*\equiv m_{n,p} + g_{\sigma} \sigma$ the effective nucleon mass. The boson fields $\sigma$, $\omega$, $\rho$, and $A$ take mean values with only the time components due to time-reversal symmetry. Then the field tensors $\omega_{\mu\nu}$, $\rho_{\mu\nu}$, and $A_{\mu\nu}$ vanish except for
\begin{equation}
\omega_{i0} = -\omega_{0i} = \partial_i \omega,
 \rho_{i0}  = -\rho_{0i}   = \partial_i  \rho,
  A_{i0}    = -A_{0i}      = \partial_i A.\nonumber
\end{equation}

The nonlinear self couplings of the mesons are determined by
\begin{equation}
U(\sigma, \omega) = -\frac{1}{3}g_2\sigma^3 - \frac{1}{4}g_3\sigma^4 + \frac{1}{4}c_3\omega^4,  \label{eq:U_NL}
\end{equation}
which effectively account for the in-medium effects and are essential for the covariant density functionals NL3~\cite{Lalazissis1997_PRC55-540}, PK1~\cite{Long2004_PRC69-034319}, TM1~\cite{Sugahara1994_NPA579-557}, GM1~\cite{Glendenning1991_PRL67-2414}, and MTVTC~\cite{Maruyama2005_PRC72-015802} adopted here.
Alternatively, the in-medium effects can be treated with density dependent coupling constants according to the Typel-Wolter ansatz~\cite{Typel1999_NPA656-331}, where
\begin{eqnarray}
g_{\xi}(n_\mathrm{b}) &=& g_{\xi} a_{\xi} \frac{1+b_{\xi}(n_\mathrm{b}/n_0+d_{\xi})^2}
                          {1+c_{\xi}(n_\mathrm{b}/n_0+d_{\xi})^2}, \label{eq:ddcp_TW} \\
g_{\rho}(n_\mathrm{b}) &=& g_{\rho} \exp{\left[-a_\rho(n_\mathrm{b}/n_0 + b_\rho)\right]}. \label{eq:ddcp_rho}
\end{eqnarray}
Here $\xi=\sigma$, $\omega$ and the baryon number density $n_\mathrm{b} = n_p+n_n$ with $n_0$ being the saturation density. In addition to the nonlinear ones, we have also adopted the density-dependent covariant density functionals DD-LZ1~\cite{Wei2020_CPC44-074107}, DDME-X~\cite{Taninah2020_PLB800-135065}, PKDD~\cite{Long2004_PRC69-034319}, DD-ME2~\cite{Lalazissis2005_PRC71-024312}, DD2~\cite{Typel2010_PRC81-015803}, and TW99~\cite{Typel1999_NPA656-331}, where the nonlinear self-couplings in Eq.~(\ref{eq:U_NL}) vanish with $g_2=g_3=c_3=0$. For completeness, the parameter sets adopted in this work are listed in Table~\ref{table:param}, where $a_{\sigma, \omega}=1$ and $b_{\sigma, \omega}=c_{\sigma, \omega}=a_\rho=0$ if nonlinear self-couplings are adopted.

\begin{table*}
\caption{\label{table:param} The adopted parameters for the covariant density functionals with nonlinear self couplings (NL3~\cite{Lalazissis1997_PRC55-540}, PK1~\cite{Long2004_PRC69-034319}, TM1~\cite{Sugahara1994_NPA579-557}, GM1~\cite{Glendenning1991_PRL67-2414}, MTVTC~\cite{Maruyama2005_PRC72-015802}) and density-dependent couplings (DD-LZ1~\cite{Wei2020_CPC44-074107}, DDME-X~\cite{Taninah2020_PLB800-135065}, PKDD~\cite{Long2004_PRC69-034319}, DD-ME2~\cite{Lalazissis2005_PRC71-024312}, DD2~\cite{Typel2010_PRC81-015803}, TW99~\cite{Typel1999_NPA656-331}).}
\begin{tabular}{c|cccccccc|ccc} \hline \hline
       & $m_n$   & $m_p$   & $m_\sigma$& $m_\omega$ & $m_\rho$ & $g_\sigma$  & $g_\omega$ & $g_\rho$  &  $g_2$      &   $g_3$   & $c_3$  \\
       &   MeV   &   MeV   &   MeV     &      MeV   &   MeV    &             &            &           & fm${}^{-1}$ &           &        \\ \hline
NL3    & 939     &   939   &  508.1941 &    782.501 &    763   &  10.2169    &  12.8675   & 4.4744    &  $-$10.4307 & $-$28.8851& 0      \\
PK1    & 938     &   938   &  511.198  &    783     &    770   &  10.0289    &  12.6139   & 4.6322    &  $-$7.2325  &    0.6183 & 71.3075\\
TM1    &939.5731 &938.2796 &  514.0891 &    784.254 &    763   &  10.3222    &  13.0131   & 4.5297    &  $-$8.1688  & $-$9.9976 & 55.636 \\
GM1    & 938     &   938   &  510      &    783     &    770   &  8.87443    &  10.60957  & 4.09772   &  $-$9.7908  & $-$6.63661& 0      \\
MTVTC  & 938     &   938   &  400      &    783     &    769   &  6.3935     &  8.7207    & 4.2696    &  $-$10.7572 & $-$4.04529& 0      \\ \hline
DD-LZ1 & 938.9   &  938.9  &538.619216 &    783     &    769   &  12.001429  &  14.292525 & 7.575467  &       0     &    0      & 0  \\
DDME-X & 938.5   &  938.5  &547.332728 &    783     &    763   &  10.706722  &  13.338846 & 3.619020  &       0     &    0      & 0  \\
PKDD   &939.5731 &938.2796 &555.5112   &    783     &    763   &  10.7385    &  13.1476   & 4.2998    &       0     &    0      & 0  \\
DD-ME2 & 938.5   &  938.5  &550.1238   &    783     &    763   &  10.5396    &  13.0189   & 3.6836    &       0     &    0      & 0  \\
DD2    &939.56536&938.27203&546.212459 &    783     &    763   &  10.686681  &  13.342362 & 3.626940  &       0     &    0      & 0  \\
TW99   & 939     &  939    &  550      &    783     &    763   &  10.7285    &  13.2902   & 3.6610    &       0     &    0      & 0  \\
\hline
\end{tabular}

\begin{tabular}{c|cccc|cccc|cc} \hline \hline
       & $a_\sigma$&$b_\sigma$&$c_\sigma$&$d_\sigma$ &$a_\omega$& $b_\omega$ & $c_\omega$ & $d_\omega$ & $a_\rho$ & $b_\rho$  \\ \hline
DD-LZ1 &1.062748 &1.763627 &2.308928 & 0.379957 &1.059181 &0.418273 &0.538663 &0.786649 &0.776095 &    0 \\
DDME-X &1.397043 &1.334964 &2.067122 & 0.401565 &1.393601 &1.019082 &1.605966 &0.455586 &0.620220 & $-$1 \\
PKDD   &1.327423 &0.435126 &0.691666 & 0.694210 &1.342170 &0.371167 &0.611397 &0.738376 &0.183305 & $-$1 \\
DD-ME2 &1.3881   &1.0943   &1.7057   & 0.4421   &1.3892   &0.9240   & 1.4620  &0.4775   &0.5647   & $-$1 \\
DD2    &1.357630 &0.634442 &1.005358 & 0.575810 &1.369718 &0.496475 &0.817753 &0.638452 &0.983955 & $-$1 \\
TW99   &1.365469 &0.226061 &0.409704 & 0.901995 &1.402488 &0.172577 &0.344293 &0.983955 &0.515000 & $-$1 \\
\hline
\end{tabular}
\end{table*}

Carrying out standard variational procedure, the equations of motion for bosons are fixed by
\begin{eqnarray}
(-\nabla^2 + m_\sigma^2) \sigma &=& -g_{\sigma} n_\mathrm{s} - g_2\sigma^2 - g_3\sigma^3, \label{eq:KG_sigma} \\
(-\nabla^2 + m_\omega^2) \omega &=& g_{\omega} n_\mathrm{b} + c_3\omega^3, \label{eq:KG_omega}\\
(-\nabla^2 + m_\rho^2) \rho     &=& \sum_{i=n,p} g_{\rho}\tau_{i} n_i, \label{eq:KG_rho}\\
                   -\nabla^2 A  &=& e(n_p - n_e - n_\mu). \label{eq:KG_photon}
\end{eqnarray}
The scalar and vector densities are determined by
\begin{eqnarray}
n_{s} &=& \sum_{i=n,p} \langle \bar{\psi}_i \psi_i \rangle = \sum_{i=n,p} \frac{{M^*}^3}{2\pi^2} f\left(\frac{\nu_i}{M^*}\right),\\
n_i &=& \langle \bar{\psi}_i\gamma^0 \psi_i \rangle = \frac{\nu_i^3}{3\pi^2},
\end{eqnarray}
where $\nu_i$ represents the Fermi momentum and $f(x) = x \sqrt{x^2+1} - \mathrm{arcsh}(x)$. The total energy of the system is then fixed by
\begin{equation}
E=\int \langle {\cal{T}}_{00} \rangle \mbox{d}^3 r, \label{eq:energy}
\end{equation}
with the energy momentum tensor
\begin{eqnarray}
\langle {\cal{T}}_{00} \rangle
&=& \sum_{i=n,p,e,\mu} \frac {{m^*_i}^4}{8\pi^{2}} \left[x_i(2x_i^2+1)\sqrt{x_i^2+1}-\mathrm{arcsh}(x_i) \right] \nonumber \\
&&   + \frac{1}{2}(\nabla \sigma)^2 + \frac{1}{2}m_\sigma^2 \sigma^2 + \frac{1}{2}(\nabla \omega)^2 + \frac{1}{2}m_\omega^2 \omega^2 + c_3\omega^4 \nonumber \\
&&   + \frac{1}{2}(\nabla \rho)^2 + \frac{1}{2}m_\rho^2 \rho^2
     + \frac{1}{2}(\nabla A)^2 - U(\sigma, \omega),
\label{eq:ener_dens}
\end{eqnarray}
where $x_i\equiv \nu_i/m^*_i$ with $m_e^*=m_e = 0.511$ MeV and $m_\mu^*=m_\mu = 105.66$ MeV.

In the Thomas-Fermi approximation (TFA), the optimum density distributions $n_i(\vec{r})$ are fixed by minimizing the total energy $E$ at given total particle numbers $N_i=\int n_i \mbox{d}^3 r$, dimension $D$, and WS cell size $R_\mathrm{W}$, which follows the constancy of chemical potentials, i.e.,
\begin{equation}
\mu_i(\vec{r}) = \sqrt{{\nu_i}^2+{m_i^*}^2} + \Sigma^\mathrm{R} + g_{\omega} \omega + g_{\rho}\tau_{i} \rho + q_i  A = \rm{constant}. \label{eq:chem_cons}
\end{equation}
Note that the ``rearrangement" term $\Sigma^\mathrm{R}$ needs to be considered if the density-dependent couplings are adopted in the Lagrangian density~\cite{Lenske1995_PLB345-355}, i.e.,
\begin{equation}
\Sigma^\mathrm{R}=
 \frac{\mbox{d} g_\sigma}{\mbox{d} n_\mathrm{b}} \sigma n_\mathrm{s}+
   \frac{\mbox{d} g_\omega}{\mbox{d} n_\mathrm{b}} \omega n_\mathrm{b}+
   \frac{\mbox{d} g_\rho}{\mbox{d} n_\mathrm{b}} \rho \sum_i\tau_i n_i.
\label{eq:re_B}
\end{equation}

\subsection{\label{sec:the_EOS} Microscopic structures of neutron star matter}

Neutron star matter at different densities exhibits various microscopic structures. At $n_\mathrm{b}\lesssim 0.0003\ \mathrm{fm}^{-3}$, neutron rich nuclei and electrons form Coulomb lattices, which can be found in the outer crusts of neutron stars and white dwarfs. At larger densities, neutrons start to drip out and form neutron gas, then the neutron star matter is essentially a liquid-gas mixed phase and can by found in the inner crust region of a neutron star. As density increases, the liquid phase will eventually take non-spherical shapes that resembles pasta, which are hence referred to as nuclear pasta~\cite{Baym1971_ApJ170-299, Negele1973_NPA207-298, Ravenhall1983_PRL50-2066, Hashimoto1984_PTP71-320, Williams1985_NPA435-844}. At densities $n_\mathrm{b}\gtrsim 0.08\ \mathrm{fm}^{-3}$, the core-crust transition takes place inside a neutron star, where the uniform phase is energetically more favorable for neutron star matter.

To obtain the microscopic structures of neutron star matter, we solve the Klein-Gordon equations and the density distributions iteratively inside a WS cell. Adopting the spherical and cylindrical approximations~\cite{Maruyama2005_PRC72-015802}, the derivatives in the Klein-Gordon equations~(\ref{eq:KG_sigma}-\ref{eq:KG_photon}) are then reduced to one-dimensional, i.e.,
\begin{eqnarray}
 \mathrm{1D:}\ \ \ \  && \nabla^2 \phi(\vec{r}) = \frac{\mbox{d}^2\phi(r)}{\mbox{d}r^2}; \label{eq:dif_1D} \\
 \mathrm{2D:}\ \ \ \  && \nabla^2 \phi(\vec{r}) = \frac{\mbox{d}^2\phi(r)}{\mbox{d}r^2} + \frac{1}{r} \frac{\mbox{d}\phi(r)}{\mbox{d}r}; \label{eq:dif_2D}\\
 \mathrm{3D:}\ \ \ \  && \nabla^2 \phi(\vec{r}) = \frac{\mbox{d}^2\phi(r)}{\mbox{d}r^2} + \frac{2}{r} \frac{\mbox{d}\phi(r)}{\mbox{d}r}, \label{eq:dif_3D}
\end{eqnarray}
which can be solved via fast cosine transformation fulfilling the reflective boundary conditions at $r=0$ and $r=R_\mathrm{W}$~\cite{Xia2021_PRC103-055812}. The density distributions of fermions are obtained with Eq.~(\ref{eq:chem_cons}) fulfilling the $\beta$-stability condition $\mu_n=\mu_p+\mu_e=\mu_p+\mu_\mu$, where in practice we have adopted the imaginary time step method~\cite{Levit1984_PLB139-147} to obtain the density profiles for the next iteration. Note that at each iteration, the total particle numbers fulfill global charge neutrality condition
\begin{equation}
  \int \left[n_p(\vec{r}) - n_e(\vec{r}) - n_\mu(\vec{r})\right] \mbox{d}^3 r\equiv 0.
\end{equation}
Different types of microscopic structures can be obtained with Eqs.~(\ref{eq:dif_1D}-\ref{eq:dif_3D}), i.e., droplet, rod, slab, tube, bubble, and uniform. At given average baryon number density $n_\mathrm{b}$, we then search for the energy minimum among six types of nuclear matter structures with optimum cell sizes $R_\mathrm{W}$. Note that the effects of charge screening are included in our calculation with electrons move freely within WS cells, which is expected to affect the microscopic structures of nuclear pasta~\cite{Maruyama2005_PRC72-015802}. With the density profiles fixed by fulfilling the convergency condition, the droplet size $R_\mathrm{d}$ and WS cell size $R_\mathrm{W}$ are then determined by
\begin{equation}
 R_\mathrm{d} =
 \left\{\begin{array}{l}
   R_\mathrm{W}\left(\frac{\langle n_p \rangle^2}{\langle n_p^2 \rangle}\right)^{1/D},  \text{\ \ \ \ \ \ \  droplet-like}\\
   R_\mathrm{W} \left(1- \frac{\langle n_p \rangle^2}{\langle n_p^2 \rangle}\right)^{1/D},  \text{\ \ bubble-like}\\
 \end{array}\right.,  \label{Eq:Rd}
\end{equation}
where $\langle n_p^2 \rangle = \int n_p^2(\vec{r}) \mbox{d}^3 r/V$ and $\langle n_p \rangle  = \int n_p(\vec{r}) \mbox{d}^3 r/V$ with the WS cell volume
\begin{equation}
  V =
 \left\{\begin{array}{l}
   \frac{4}{3}\pi R_\mathrm{W}^3,\  D = 3\\
   \pi a R_\mathrm{W}^2 , \  D = 2\\
   a^2 R_\mathrm{W}, \ \  D = 1\\
 \end{array}\right.. \label{Eq:V}
\end{equation}
In order for the volume to be finite for the slabs and rods/tubes at $D = 1$ and 2, here we have adopted a finite cell size $a = 30$ fm. Meanwhile, as we decrease the density, it is found that $R_\mathrm{W}$ grows drastically and quickly exceeds the limit for any viable numerical simulations. In such cases, as was done in our previous study~\cite{Xia2022_PRC105-045803}, at densities $n_\mathrm{b}\lesssim 10^{-4}$ fm${}^{-3}$ we divide the WS cell into a core with radius $R_\mathrm{in}=35.84$ fm and a spherical shell with constant densities.

\section{\label{sec:results} Results and Discussion}
\subsection{\label{sec:pasta_beta} Neutron star matter}

\begin{table}
\caption{\label{table:NM} Saturation properties of nuclear matter corresponding to the covariant density functionals indicated in Table~\ref{table:param}. }
\begin{tabular}{c|ccccccc} \hline \hline
       & $n_0$        &   $B$    &   $K$  &  $J$   & $S$    &  $L$  & $K_\mathrm{sym}$        \\
       & fm${}^{-3}$  &   MeV    &   MeV  &  MeV   &  MeV   &  MeV  &   MeV             \\ \hline
NL3    &  0.148       & $-$16.25 &  271.7 &  204   & 37.4   & 118.6 &   101           \\
PK1    &  0.148       & $-$16.27 &  282.7 &$-27.8$ & 37.6   & 115.9 &   55           \\
TM1    &  0.145       & $-$16.26 &  281.2 & $-285$ & 36.9   & 110.8 &   34           \\
GM1    &  0.153       & $-$16.33 &  300.5 & $-216$ & 32.5   &  94.0 &   18           \\
MTVTC  &  0.153       & $-$16.30 &  239.8 & $-513$ & 32.5   &  89.6 &  $-6.5$         \\  \hline
DD-LZ1 &  0.158       & $-$16.06 &  230.7 & 1330   & 32.0   &  42.5 &  $-20$            \\
DDME-X &  0.152       & $-$16.11 &  267.6 & 874    & 32.3   &  49.7 &  $-72$           \\
PKDD   &  0.150       & $-$16.27 &  262.2 & $-119$ & 36.8   &  90.2 &  $-81$           \\
DD-ME2 &  0.152       & $-$16.13 &  250.8 & 477    & 32.3   &  51.2 &  $-87$           \\
DD2    &  0.149       & $-$16.02 &  242.7 & 169    & 31.7   &  55.0 &  $-93$          \\
TW99   &  0.153       & $-$16.24 &  240.2 & $-540$ & 32.8   &  55.3 &  $-125$          \\
\hline
\end{tabular}
\end{table}

The nuclear matter properties around the saturation density are illustrated in Table~\ref{table:NM} for various covariant density functionals adopted here, which covers a wide range for the incompressibility $K$, the skewness coefficient $J$, the symmetry energy $S$, the slope $L$ and curvature parameter $K_\mathrm{sym}$ of nuclear symmetry energy. Based on those functionals, we then investigate the EOSs and microscopic structures of neutron star matter adopting the numerical recipe introduced in Sec.~\ref{sec:the}.

\begin{figure*}
\centering
\includegraphics[width=0.7\linewidth]{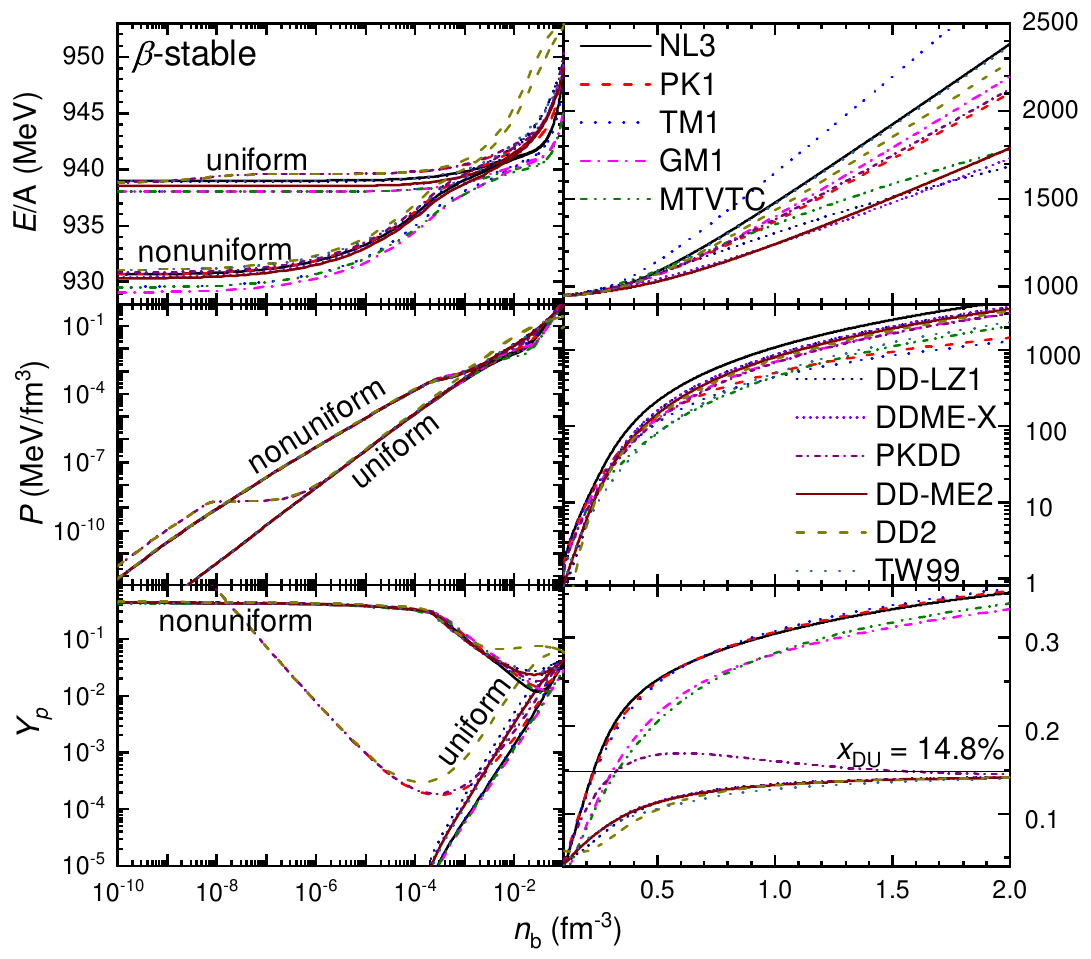}
\caption{\label{Fig:EOS-beta} The energy per baryon $E/A$, pressure $P$, and proton fraction $Y_p$ of neutron star matter as functions of baryon number density $n_\mathrm{b}$, which are obtained with various covariant density functionals indicated in Table~\ref{table:param}. Both the uniform and nonuniform phases are illustrated in the left panels, while only uniform phases are presented in the right panels since the nonuniform one does not emerge.}
\end{figure*}

In Fig.~\ref{Fig:EOS-beta} we present the obtained energy per baryon, pressure, and proton fraction for the most favorable nuclear shape with optimum WS cell size $R_\mathrm{W}$. For comparison, the corresponding results for uniform matter are presented in the left panels as well. As we decrease the density, the proton fraction $Y_p$ of the uniform phase decreases and eventually vanishes for most of the functionals. Nevertheless, as indicated in Table~\ref{table:param}, adopting realistic neutron and proton masses for the covariant density functionals TM1, PKDD, and DD2, the proton fraction $Y_p$ of the uniform phase does not vanish but increases to 1 as we decrease the density at $n_\mathrm{b} \lesssim 10^{-4}$ fm${}^{-3}$, which is reasonable as protons are more stable than neutrons. The contribution of electrons are then present in order to reach local charge neutrality condition $n_p=n_e$. Once nonuniform nuclear structures emerge, the proton fraction $Y_p$ deviates significantly from that of the uniform phase, which approaches to $Y_p=0.43$-0.45 at vanishing densities. The energy per baryon are then reduced by up to 8 MeV. Note that the absolute values of the energy per baryon at vanishing densities are sensitive to the adopted nucleon masses, while the obtained binding energy for various functionals coincide with each other.

At vanishing densities, the pressure mainly comes from the contributions of electrons and is thus increasing with $Y_p$. Except for those adopting realistic nucleon masses, the obtained pressure for the nonuniform phase is larger than that of the uniform one as predicted by most of the functionals. The neutron drip densities $n_\mathrm{d}$ can be obtained by equating the chemical potential of neutrons with their mass, i.e., $\mu_n(n_\mathrm{d})=m_n$. The obtained values of $n_\mathrm{d}$ for various functionals are then indicated in Table~\ref{table:phase}, where those with the density-dependent couplings generally predict smaller neutron drip densities compared with that of nonlinear ones. Then at $n_\mathrm{b} \lesssim 10^{-4}$ fm${}^{-3}<n_\mathrm{d}$, neutron star matter are comprised of Coulomb lattices of nuclei and electrons, where the similar values for the pressure are obtained with various functionals in this density range. In such cases, the EOSs of neutron star matter at $n_\mathrm{b} \lesssim 10^{-4}$ fm${}^{-3}$ generally coincide with each other except for the slight differences (within 0.1\%) in the energy density due to the variations in the nucleon masses indicated in Table~\ref{table:param}.

\begin{figure*}[htbp]
\begin{minipage}[t]{0.49\linewidth}
\centering
\includegraphics[width=\textwidth]{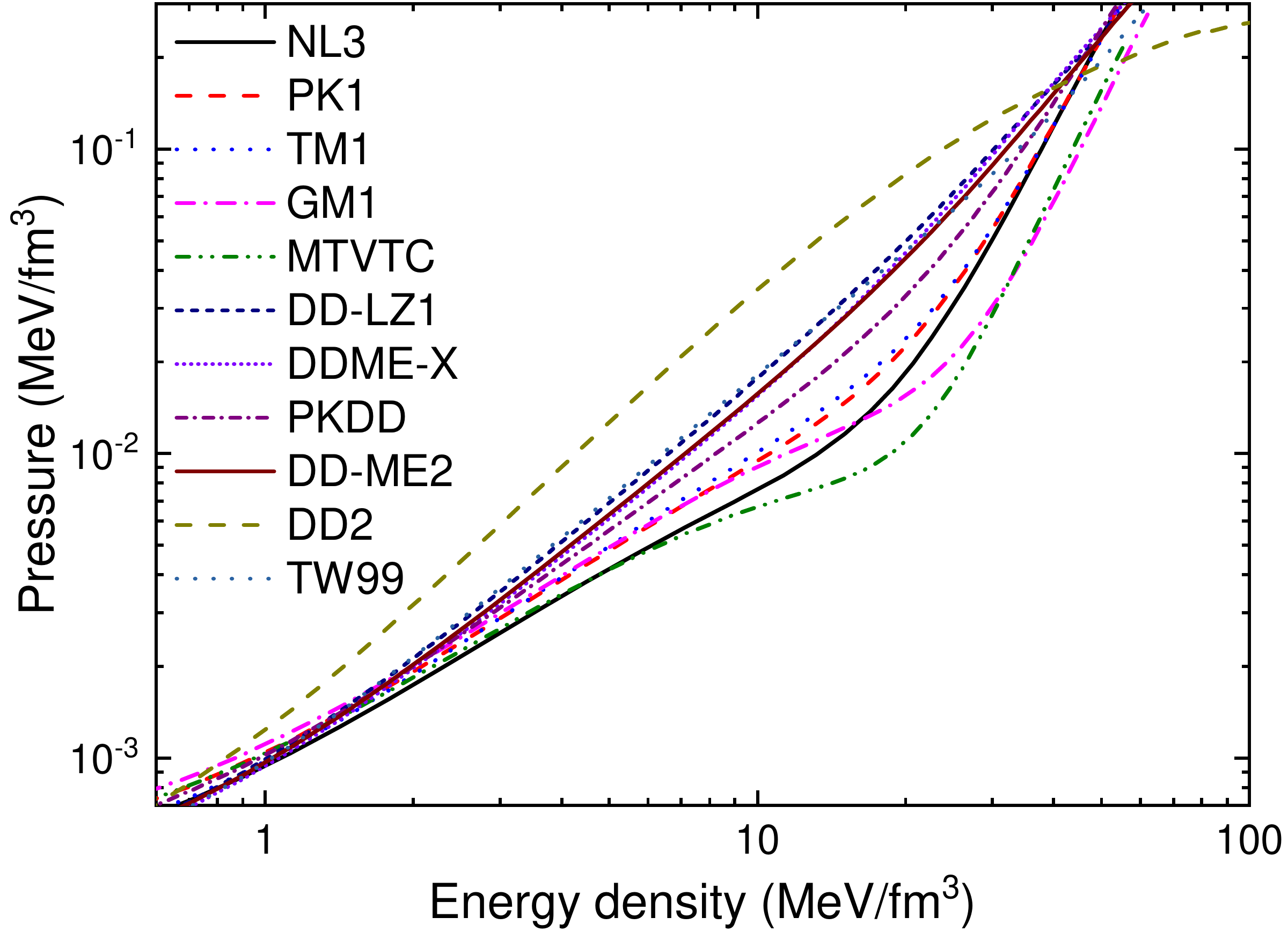}
\end{minipage}%
\hfill
\begin{minipage}[t]{0.49\linewidth}
\centering
\includegraphics[width=\textwidth]{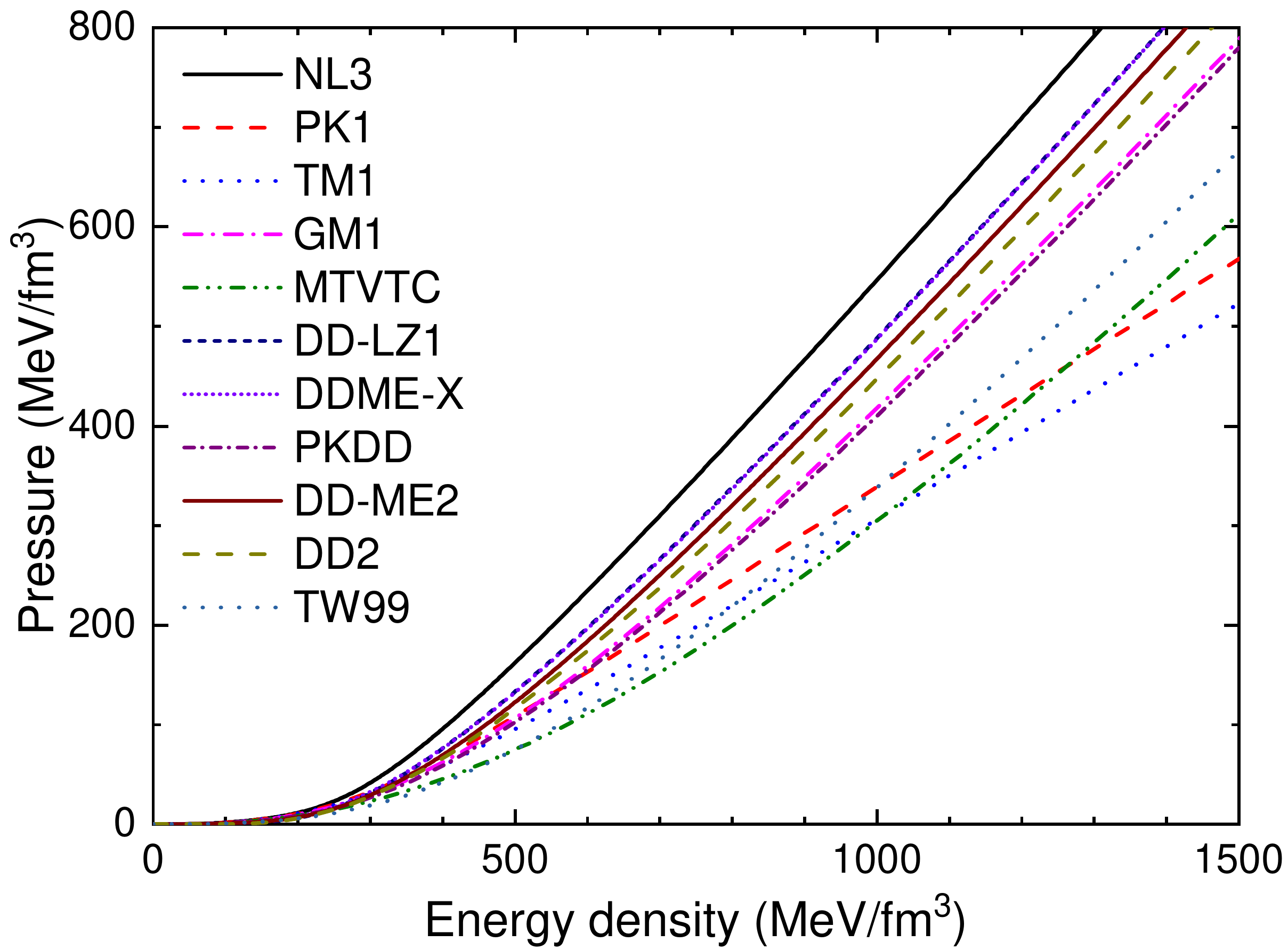}
\end{minipage}
\caption{\label{Fig:EOS-all} The EOSs for the pasta (left) and uniform (right) phases of neutron star matter, in correspondence to Fig.~\ref{Fig:EOS-beta}.}
\end{figure*}

At larger densities with $n_\mathrm{b} \gtrsim n_\mathrm{d}$, we note that the slope of the energy per baryon, pressure, and proton fraction change suddenly as neutron gas starts to coexist with the liquid phase of nuclear matter, which forms the nuclear pasta typically found in the inner crusts of neutron stars. In contrast to the stellar matter located in the outer crusts of neutron stars, as indicated in the left panel of Fig.~\ref{Fig:EOS-all}, the EOSs of the pasta phase are sensitive to the adopted nuclear energy density functional. It is found that the EOSs obtained with nonlinear couplings become stiffer at the energy density $E/n_\mathrm{b} \gtrsim 20$ MeV fm${}^{-3}$ ($n_\mathrm{b} \gtrsim 0.02$ fm${}^{-3}$), while the EOSs obtained with density-dependent couplings vary more smoothly with density. In general, the relative uncertainty in the EOSs of the pasta phase grows with density and then decreases once reaches the peak at $E/n_\mathrm{b} \approx 20$ MeV fm${}^{-3}$. The corresponding differences in the EOSs at subsaturation densities are expected to affect the radii and crust thickness of neutron stars, which will be illustrated in Fig.~\ref{Fig:MR}. Note that for the functional DD2, the obtained results for nuclear pasta deviate significantly from other functionals. This is mainly because we have employed the single nucleus approximation (SNA) and neglected the contributions of light clusters as initially proposed in Ref.~\cite{Typel2010_PRC81-015803}. For more suitable treatments adopting the extended nuclear statistical equilibrium model, one can refer to Ref.~\cite{Fischer2014_EPJA50-46} with the publicly available EOS HS(DD2), which is more reasonable than the DD2 EOS presented in the left panel of Fig.~\ref{Fig:EOS-all} with a too large proton fraction.

\begin{table*}
\caption{\label{table:phase} Densities (in fm${}^{-3}$) for shape transitions, which are obtained by varying the density in a step of 0.002 fm${}^{-3}$. The neutron drip densities $n_\mathrm{d}$ obtained with $\mu_n(n_\mathrm{d}) = m_n$ and critical densities $n_\mathrm{DU}$ for the occurrence of DU processes with $Y_p(n_\mathrm{DU})=14.8\%$ are indicated as well.}
\begin{tabular}{c|ccccc|cccccc} \hline \hline
 Transition              &  NL3  &  PK1  &  TM1  &  GM1  & MTVTC & DD-LZ1 & DDME-X & PKDD  & DD-ME2 & DD2   & TW99    \\ \hline
$n_\mathrm{d}$ ($10^{-4}$) &  2.4  &  2.7  &  2.3  & 3.1   &  3.1  &  1.9   & 1.9    &  2.3  & 2.0    & 1.7   & 1.8    \\ \hline
droplet-rod              &   -   &   -   &   -   &   -   &   -   & 0.059  & 0.065  &   -   & 0.063  & 0.041 & 0.063  \\
rod-slab                 &   -   &   -   &   -   &   -   &   -   & 0.065  & 0.073  &   -   & 0.071  & 0.057 & 0.071  \\
slab-tube                &   -   &   -   &   -   &   -   &   -   & 0.069  &   -    &   -   & 0.073  & 0.087 & 0.075  \\
tube-bubble              &   -   &   -   &   -   &   -   &   -   &   -    &   -    &   -   &   -    & 0.101 &   -    \\
core-crust               & 0.057 & 0.061 & 0.061 & 0.067 & 0.061 & 0.071  & 0.077  & 0.065 & 0.075  & 0.111 & 0.077  \\ \hline
$n_\mathrm{DU}$          & 0.228 & 0.230 & 0.236 & 0.309 & 0.328 &   -    &   -    & 0.325 &   -    &   -   &   -    \\ \hline
\end{tabular}
\end{table*}

If we further increase the density, the uniform phase becomes energetically more favorable once exceeding the core-crust transition densities indicated in Table~\ref{table:phase}, e.g., $n_\mathrm{b} \gtrsim 0.08$ fm${}^{-3}$. The corresponding energy per baryon, pressure, and proton fraction of neutron star matter are indicated in the right panels of Fig.~\ref{Fig:EOS-beta}. In contrast to the cases at smaller densities, the uncertainties in those quantities grow drastically as density increases at $n_\mathrm{b} \gtrsim 0.3$ fm${}^{-3}$, where the less constrained higher order coefficients such as $L$ and $K_\mathrm{sym}$ start to play important roles. If the EOSs do not cross with each other, we note that the stiffness of EOS is directly linked to the maximum mass of neutron stars as indicated in Fig.~\ref{Fig:MR}. Despite their evident differences in the energy per baryon, as indicated in the right panel of Fig.~\ref{Fig:EOS-all}, we note that the EOSs obtained with the functionals DD-LZ1 and DDME-X coincide with each other at $n_\mathrm{b} \gtrsim 0.08$ fm${}^{-3}$.

The obtained proton fractions of neutron star matter at $n_\mathrm{b} \gtrsim 0.08$ fm${}^{-3}$ show distinctive trends between the functionals with nonlinear self-couplings and density dependent ones, which is attributed to the differences in the higher order coefficients $L$ and $K_\mathrm{sym}$ of nuclear symmetry energy as indicated in Table~\ref{table:NM}. It is found that $Y_p$ increases with density if nonlinear self-couplings are adopted, while for density dependent ones $Y_p$ approaches to a constant value ($\sim$0.14). It is worth mentioning that if isovector scalar channel ($\delta$ meson) are included in density-dependent covariant density functionals, the proton fraction may deviation from the trend and increase with density~\cite{Wang2014_PRC90-055801}. Meanwhile, we note that a peculiar density-dependent behavior of $Y_p$ (reaching its maximum at $n_\mathrm{b} \approx 0.5$ fm${}^{-3}$) is obtained with the functional PKDD, which is attributed to the large slope $L$ but negative curvature parameter $K_\mathrm{sym}$ of nuclear symmetry energy. In principle, the proton fraction is directly connected to the most efficient cooling mechanism of neutron stars. Once the momentum conservation is fulfilled with $Y_p\gtrsim 14.8\%$, the direct Urca (DU) processes $n\rightarrow p + e^-+\bar{\nu}_e$ and $p + e^-\rightarrow n+\nu_e$ will take place and rapidly cools the neutron star down~\cite{Klaehn2006_PRC74-035802, Page2006_NPA777-497}. As indicated in Fig.~\ref{Fig:EOS-beta}, the critical densities $n_\mathrm{DU}$ for the occurrence of DU processes can be obtained once $Y_p>14.8\%$, where the corresponding values are presented in Table~\ref{table:phase}. It is found that the DU processes only take places if the functionals with nonlinear self-couplings and PKDD are employed.

\begin{figure*}[!ht]
  \centering
  \includegraphics[width=0.7\linewidth]{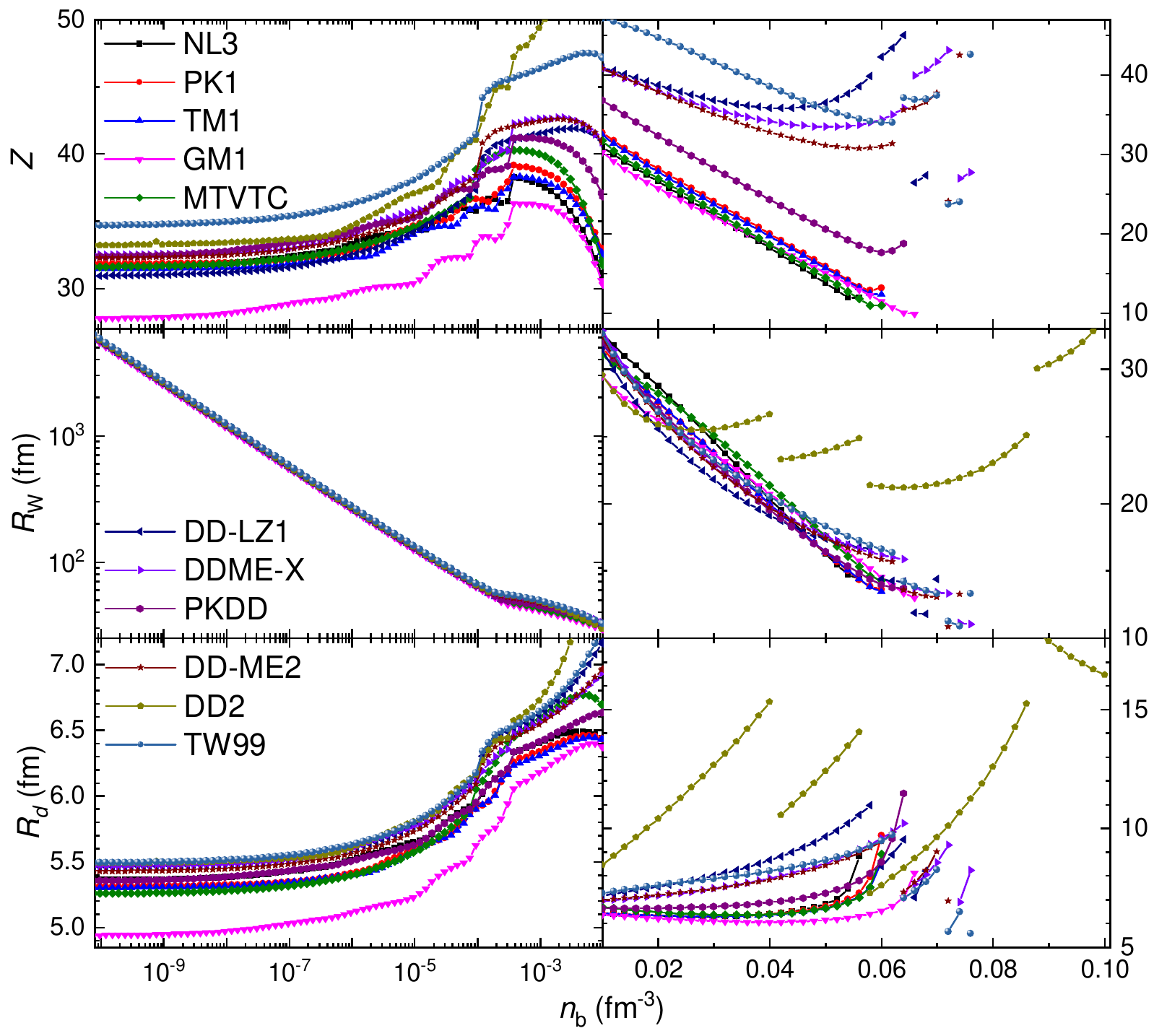}
  \caption{\label{Fig:Micro-beta}Proton number $Z$, WS cell radius $R_\mathrm{W}$, and droplet size $R_\mathrm{d}$ for the nonuniform nuclear matter typically found in neutron star crusts, where the corresponding EOSs are indicated in Fig.~\ref{Fig:EOS-beta}.}
\end{figure*}

\begin{figure}
\includegraphics[width=\linewidth]{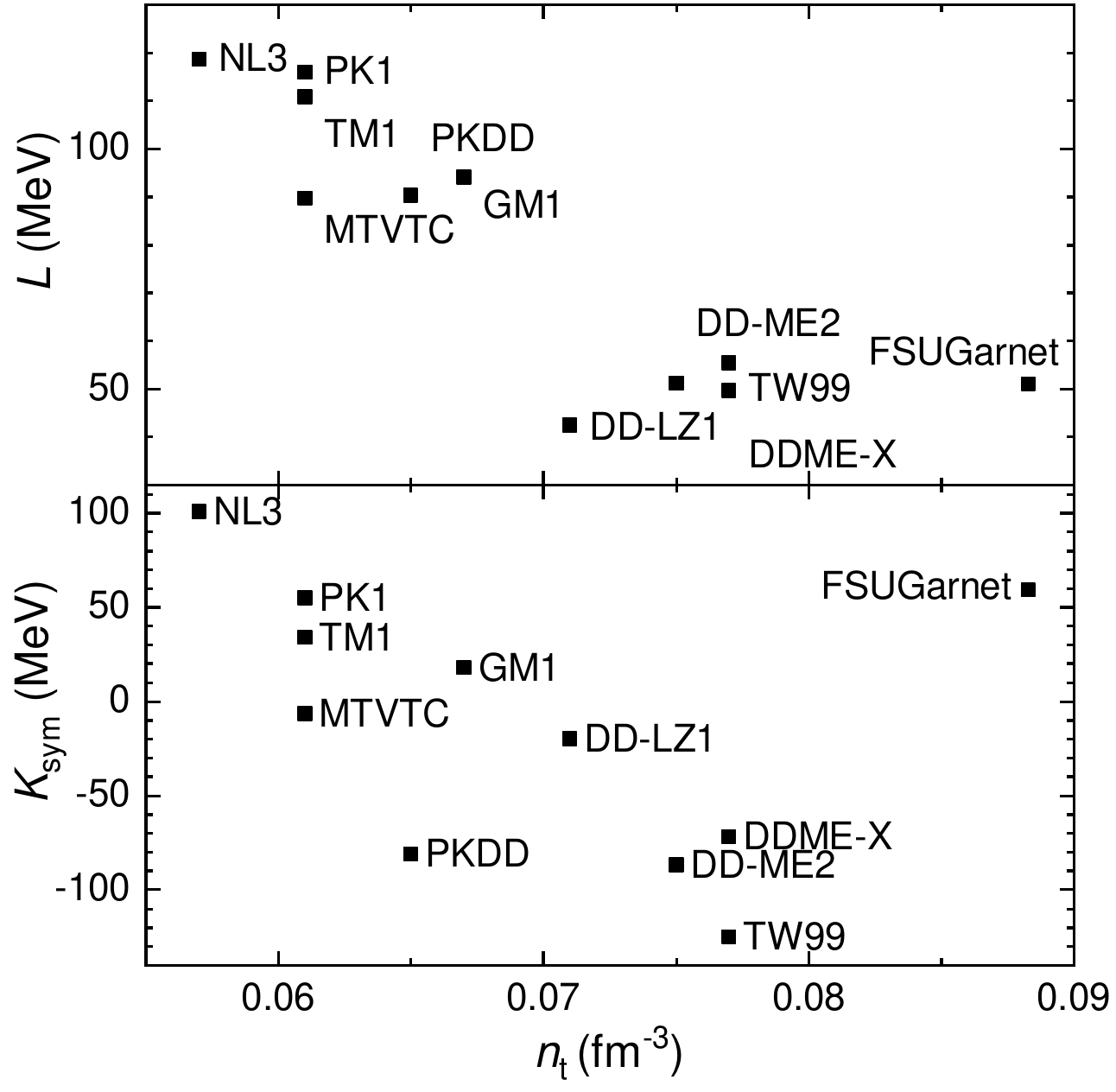}
\caption{\label{Fig:nt-sym} Crust-core transition densities $n_\mathrm{t}$ (indicated in Table~\ref{table:phase}) obtained with various covariant density functionals and the corresponding slope $L$ and curvature parameter $K_\mathrm{sym}$ of symmetry energy (indicated in Table~\ref{table:NM}). }
\end{figure}

Beside the EOSs, the variation in the microscopic structures has significant implications on the transport and elastic properties of neutron star matter, which would in turn affect various phenomenons observed in neutron stars~\cite{Chamel2008_LRR11-10, Caplan2017_RMP89-041002}. In Fig.~\ref{Fig:Micro-beta} we present the obtained microscopic structures of nonuniform neutron star matter corresponding to the EOSs in Figs.~\ref{Fig:EOS-beta} and \ref{Fig:EOS-all}, where the proton number $Z$, WS cell radius $R_\mathrm{W}$, and droplet size $R_\mathrm{d}$ are indicated. As density increases, the droplet, rod, slab, tube, bubble, and uniform phases appear sequentially for the nuclear pasta in neutron stars. The transition densities between different nuclear shapes are indicated in Table~\ref{table:phase}. We note that for the functionals predicting large slope $L$ of symmetry energy, only the droplet phase emerge for the nuclear pasta in $\beta$-equilibrium, while as indicated in Fig.~\ref{Fig:nt-sym} the core-crust transition densities $n_\mathrm{t}$ are smaller than those predicting smaller $L$ as well. This is consistent with previous studies, where the proton number of nuclei, the core-crust transition density, and the onset density of non-spherical nuclei generally decrease with $L$~\cite{Oyamatsu2007_PRC75-015801, Xu2009_ApJ697-1549, Grill2012_PRC85-055808, Bao2015_PRC91-015807, Shen2020_ApJ891-148, Xia2021_PRC103-055812}. The obtained results with the functional NL3 and DD-ME2 generally coincide with those in Ref.~\cite{Grill2012_PRC85-055808} with slightly larger core-crust transition density, while those of TM1 coincide with Ref.~\cite{Bao2015_PRC91-015807}. Meanwhile, according to Fig.~\ref{Fig:nt-sym}, it is evident that $n_\mathrm{t}$ also decreases with the curvature parameter $K_\mathrm{sym}$ of symmetry energy, which is closely related to the curvature-slope correlations~\cite{Pais2012_PRL109-151101, Li2020_PRC102-045807}. To show the consequences of adopting a functional that does not follow the curvature-slope correlation, in Fig.~\ref{Fig:nt-sym} we present the results predicted by FSUGarnet using the compressible liquid drop model~\cite{Parmar2022_PRD105-043017}, where the $L$-$n_\mathrm{t}$ correlation still holds approximately but not for the $K_\mathrm{sym}$-$n_\mathrm{t}$ correlation. Note that the functional DD2 predicts rather large $n_\mathrm{t}$, which will be reduced if the extended nuclear statistical equilibrium model is adopted including the contributions of light clusters~\cite{Fischer2014_EPJA50-46}.

Similar to the EOSs of neutron star matter, as indicated in Fig.~\ref{Fig:Micro-beta}, the microscopic structures vary little with respect to the adopted functionals at $n_\mathrm{b} \lesssim 10^{-4}$ fm${}^{-3}$. For example, slightly different proton numbers and droplet sizes are obtained at vanishing densities (e.g., $n_\mathrm{b} \approx 10^{-10}$ fm) adopting various functionals, which vary within the ranges $Z\approx28$-35 and $R_\mathrm{d}\approx 5$-5.5 fm and increase with density at $n_\mathrm{b}\lesssim 10^{-4}$ fm${}^{-3}$. The obtained WS cell radius $R_\mathrm{W}$ is decreasing with density and is insensitive to the adopted functionals at $n_\mathrm{b} \lesssim 10^{-4}$ fm${}^{-3}$. We note that the differences on the microscopic structures start to grow once $n_\mathrm{b} \gtrsim n_\mathrm{d}$, where the values of $Z$ and $R_\mathrm{d}$ as functions of density may exhibit different trends for various functionals. At larger densities with $n_\mathrm{b} \gtrsim 0.01$ fm${}^{-3}$, the proton number $Z$ and WS cell radius $R_\mathrm{W}$ are generally decreasing, while the droplet size $R_\mathrm{d}$ increases. Throughout the vast density range considered here, in consistent with previous investigations~\cite{Oyamatsu2007_PRC75-015801, Xu2009_ApJ697-1549, Grill2012_PRC85-055808, Bao2015_PRC91-015807, Shen2020_ApJ891-148, Xia2021_PRC103-055812}, the obtained values of $Z$ and $R_\mathrm{d}$ approximately decrease with $L$ if different functionals are adopted, while the values of $R_\mathrm{W}$ are close to each other. Note that rather large values of $Z$, $R_\mathrm{d}$, and $R_\mathrm{W}$ are obtained with the functional DD2, which is mainly due to the SNA adopted here and neglecting light clusters.

\subsection{\label{sec:star} Neutron stars}
\begin{figure}
\includegraphics[width=\linewidth]{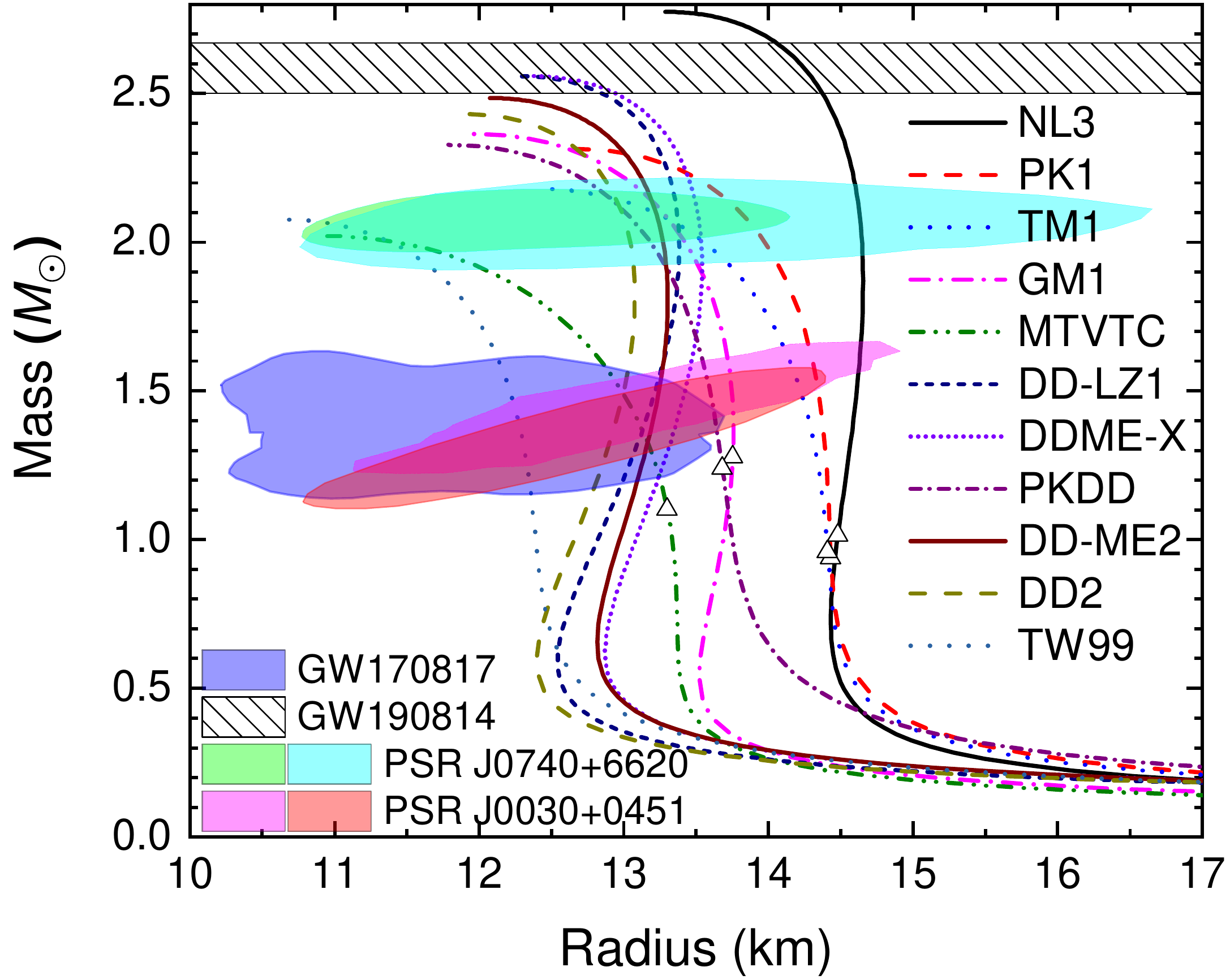}
\caption{\label{Fig:MR} Mass-radius relations of neutron stars corresponding to the EOSs presented in Figs.~\ref{Fig:EOS-beta} and \ref{Fig:EOS-all}, while the open triangles are the critical masses $M_\mathrm{DU}$ for DU processes. The shaded regions indicate the constraints from the binary neutron star merger event GW170817 within 90\% credible region~\cite{LVC2018_PRL121-161101}, the observational pulse-profiles in PSR J0030+0451 and PSR J0740+6620 within 68\% credible region~\cite{Riley2019_ApJ887-L21, Riley2021_ApJ918-L27, Miller2019_ApJ887-L24, Miller2021_ApJ918-L28}, and the mass (2.50-2.67 $M_{\odot}$) of a compact object observed in the gravitational-wave signal GW190814 in 90\% credible region~\cite{Abbott2020_ApJ896-L44}.}
\end{figure}

Based on the unified EOSs of neutron star matter presented in Figs.~\ref{Fig:EOS-beta} and \ref{Fig:EOS-all}, the structures of neutron stars are obtained by solving the TOV equation
\begin{eqnarray}
&&\frac{\mbox{d}P}{\mbox{d}r} = -\frac{G M E}{r^2}   \frac{(1+P/E)(1+4\pi r^3 P/M)} {1-2G M/r},  \label{eq:TOV}\\
&&\frac{\mbox{d}M}{\mbox{d}r} = 4\pi E r^2, \label{eq:m_star}
\end{eqnarray}
where $G=6.707\times 10^{-45}\ \mathrm{MeV}^{-2}$ is the gravity constant. In Fig.~\ref{Fig:MR} we present the mass-radius relations of neutron stars corresponding to the covariant density functionals indicated in Table~\ref{table:param}. Various constraints from pulsar observations are indicated in Fig.~\ref{Fig:MR}, i.e., the binary neutron star merger event GW170817~\cite{LVC2018_PRL121-161101}, the simultaneous measurements of masses and radii for PSR J0030+0451 and PSR J0740+6620~\cite{Riley2019_ApJ887-L21, Riley2021_ApJ918-L27, Miller2019_ApJ887-L24, Miller2021_ApJ918-L28}, and the measured mass of a compact object involved in a compact binary coalescence from the gravitational-wave signal GW190814~\cite{Abbott2020_ApJ896-L44}. The open triangles in Fig.~\ref{Fig:MR} correspond to the critical masses $M_\mathrm{DU}$ for DU processes with the central densities exceeding $n_\mathrm{DU}$. It is expected that the neutrino emissivity is enhanced significantly for neutron stars with $M>M_\mathrm{DU}$~\cite{Spinella2018_Universe4-64}, which cool down too rapidly within just a few years~\cite{Blaschke2004_AA424-979}. It is found that the DU processes only take places if the functionals with nonlinear self-couplings and PKDD are employed ($M_\mathrm{DU}\approx 0.9$-1.3 $M_{\odot}$), which have large slopes of symmetry energy with $L\gtrsim 90$ MeV.

We note that all functionals predict neutron stars with maximum masses exceeding 2 $M_{\odot}$~\cite{Fonseca2021_ApJ915-L12}, while the functionals NL3, DD-LZ1, and DDME-X predict even larger maximum masses supporting the possibility that the secondary object observed in GW190814 is a neutron star~\cite{Abbott2020_ApJ896-L44}. Nevertheless, as indicated in Table~\ref{table:NM}, the incompressibility, symmetry energy and its slope for nuclear matter obtained with the functional NL3 exceed the constraints from start-of-art studies~\cite{Shlomo2006_EPJA30-23, Zhang2020_PRC101-034303, Essick2021_PRL127-192701}, leading to neutron stars with too large radii and masses. A combined constraint on the masses and radii of neutron stars suggest that DD2, DD-LZ1, DD-ME2, and DDME-X are the most probable functionals that are consistent with observations. However, to support massive neutron stars, their skewness coefficients $J$ are much larger than expected, which was constrained to be $J=-700\pm 500$ MeV from fits of generalized Skyrme force to breathing-mode energies~\cite{Farine1997_NPA615-135} and $J=-390^{+60}_{-70}$ MeV from empirical pressures in relativistic heavy-ion collisions~\cite{Xie2021_JPG48-025110}. The maximum masses of neutron stars obtained by the two functionals MTVTC and TW99 are close to 2 $M_{\odot}$, while the corresponding radii are slightly small and locate in the lower ends of the PSR J0740+6620 constraints~\cite{Riley2021_ApJ918-L27, Miller2021_ApJ918-L28}. The functionals PKDD, GM1, TM1, PK1, and NL3 predict slightly too large radii according to the constraint derived from the binary neutron star merger event GW170817~\cite{LVC2018_PRL121-161101}, which are attributed to the much larger values for $K$ and/or $L$ as indicated in Table~\ref{table:NM}. Nevertheless, if exotic phases with the emergence of new degrees of freedom such as mesons ($\pi$, $K$, etc.), heavy baryons ($\Delta$, $\Lambda$, $\Sigma$, $\Xi$, $\Omega$, etc.), and deconfinement phase transition into quarks ($u$, $d$, $s$) were to take place, the corresponding EOSs would become softer, which effectively reduce the radii of compact stars and comply with the observational constraints~\cite{Baym2018_RPP81-056902, Sun2019_PRD99-023004, Xia2020_PRD102-023031, LI2020, Dexheimer2021_PRC103-025808, Dexheimer2021_EPJA57-216, Sun2021_PRD103-103003, Tu2022_ApJ925-16, Sun2022}.

It was augured that the sudden spin-ups (glitches) of pulsars are due to the angular momentum transfers from the superfluid component of a neutron star's interior to its solid crust~\cite{Anderson1975_Nature256-25}, whose characteristic properties could provide additional constraints on neutron star structures. In particular, the fractional crustal moment of inertia ${I_\mathrm{c}}/{I}$ can be measured with
\begin{equation}
 \frac{I_\mathrm{c}}{I} \gtrsim  \frac{2\tau_c}{T}\sum_i\left(\frac{\Delta\Omega_p}{\Omega_p} \right)_i, \label{eq:DI_tau}
\end{equation}
where $\tau_c$ represents the characteristic age of the pulsar, $T$ the total time span for glitch monitoring, and ${\Delta\Omega_p}/{\Omega_p}$ the fractional frequency jump of glitches. To explain the glitches observed in the Vela pulsar, the fractional crustal moment of inertia was constrained to be ${I_\mathrm{c}/I} \gtrsim 1.4\%$~\cite{Link1999_PRL83-3362}. However, it was argued that the entrainment of superfluid neutrons by the solid crust could lower its mobility and increase the lower limit to ${I_\mathrm{c}/I} \gtrsim 7\%$, causing the ``glitch crisis" where many nuclear EOSs fail to meet the constraint~\cite{Andersson2012_PRL109-241103, Chamel2012_PRC85-035801, Li2016_ApJS223-16}. Nevertheless, it is worth mentioning that the entrainment effect may be suppressed if the pairing gap is of order or greater than the strength of the lattice potential~\cite{Watanabe2017_PRL119-062701}, where the constraint can be reduced to ${I_\mathrm{c}/I} \gtrsim 2.4\pm0.1\%$~\cite{Li2017_IAUS13-360}.

\begin{figure}
\includegraphics[width=\linewidth]{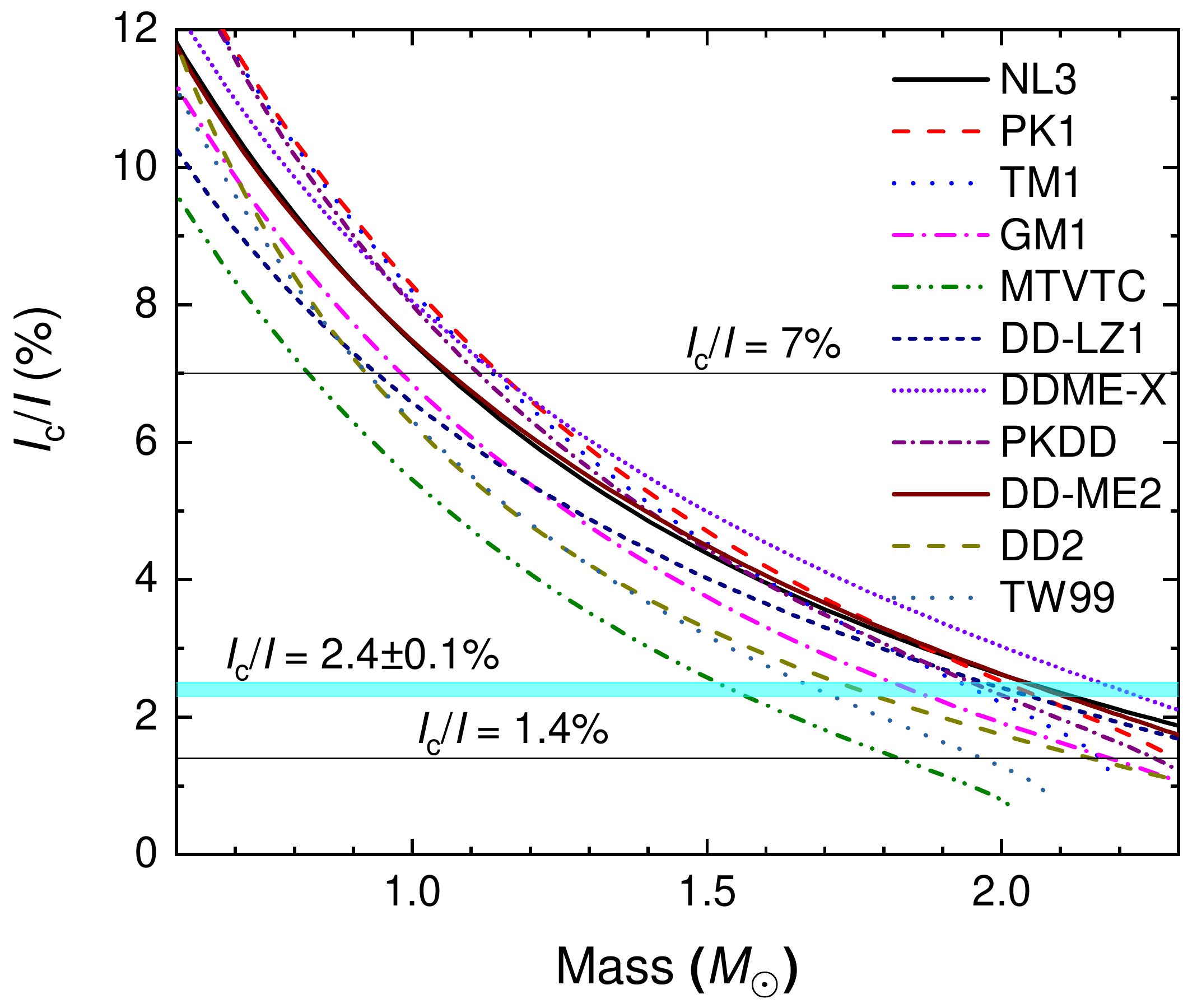}
\caption{\label{Fig:DI} Crustal fraction of moment of inertia as a function of mass for neutron stars indicated in Fig.~\ref{Fig:MR}. The horizontal lines and band represent a possible constraint derived from the glitch activities in the Vela pulsar~\cite{Link1999_PRL83-3362, Andersson2012_PRL109-241103, Chamel2012_PRC85-035801, Li2016_ApJS223-16, Watanabe2017_PRL119-062701, Li2017_IAUS13-360}.}
\end{figure}

For slowly rotating neutron stars, the fraction of crustal moment of inertia can be estimated with~\cite{Link1999_PRL83-3362}
\begin{equation}
 \frac{I_\mathrm{c}}{I} \approx \frac{28\pi P_\mathrm{t} R^3}{3M}\frac{1 - 1.67\beta -0.6\beta^2}{\beta+ \frac{2 P_\mathrm{t}}{n_\mathrm{t}m_n}\left(\frac{1}{\beta} +5-14\beta  \right)},  \label{eq:DI}
\end{equation}
where $P_\mathrm{t}$ is the pressure at core-crust transition density $n_\mathrm{t}$ and  $\beta=GM/R$ the compactness of a neutron star. The obtained results are then presented in Fig.~\ref{Fig:DI}, where the crustal moment of inertia is decreasing with mass. It is evident that ${I_\mathrm{c}/I}$ is sensitive to the EOS, and in particular the crust one since it determines the mass and thickness of a neutron star's crust. Therefore a unified treatment for the EOSs of uniform (core) and nonuniform (crust) neutron star matter is essential to obtain accurately the radii, crust properties, core-crust transition density, as well as the corresponding microscopic structures. In order to meet the constraints of Vela pulsar as indicated by the  horizontal lines and band, we note that a neutron star should not be more massive than a critical value, which varies with the EOSs and the effectiveness of the entrainment effect. Nevertheless, to distinguish the EOSs from one another, more detailed investigations on pulsar glitches are required in future studies.

\begin{figure*}
  \centering
  \includegraphics[width=0.6\linewidth]{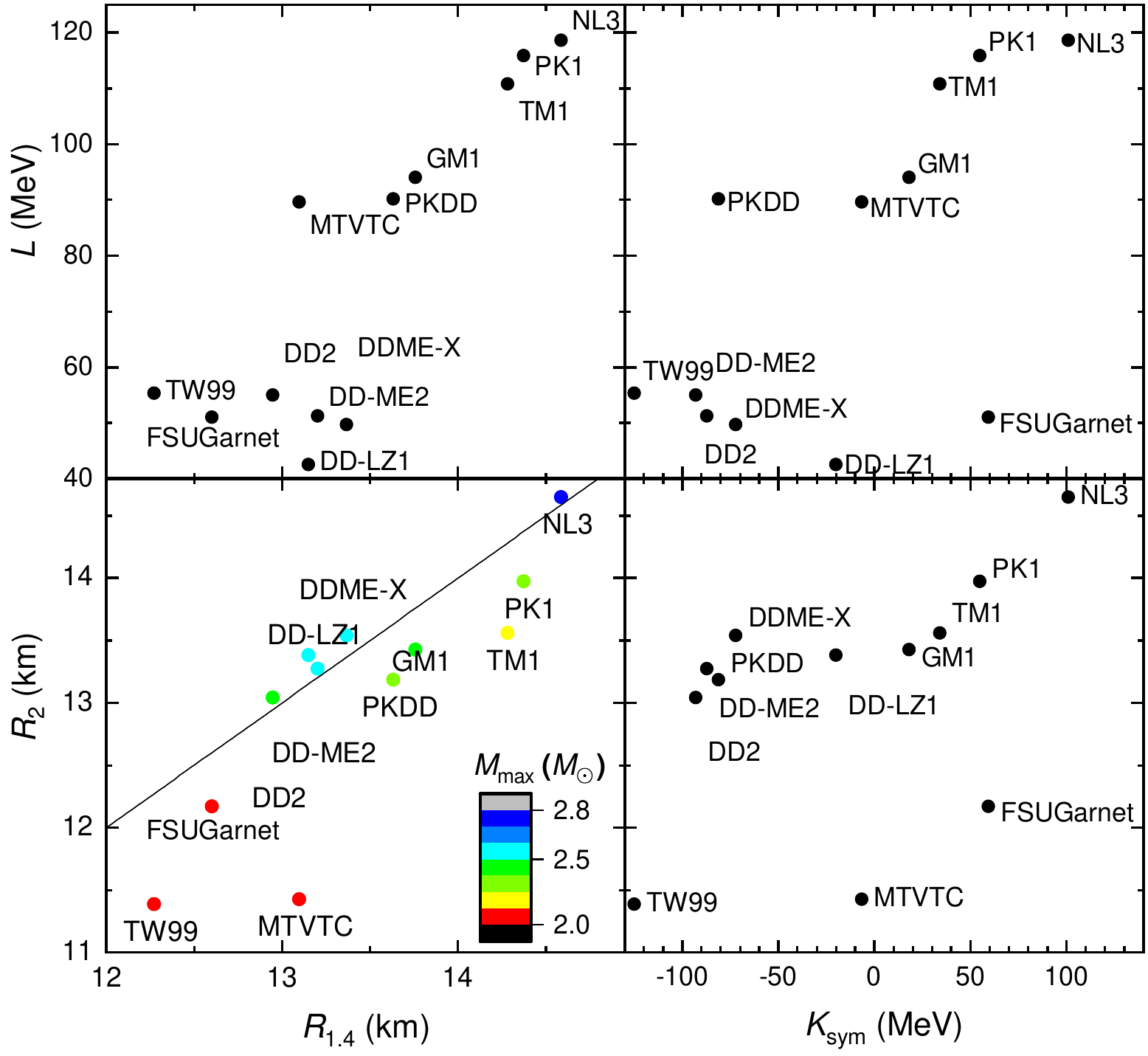}
  \caption{\label{Fig:param} Correlations between neutron stars' radii ($R_{1.4}$ at $M=1.4 M_{\odot}$ and $R_2$ at $M=2 M_{\odot}$), the slope $L$ and curvature parameter $K_\mathrm{sym}$ of nuclear symmetry energy obtained with various covariant density functionals.}
\end{figure*}

To examine the possible correlations between the macroscopic neutron star structures and microscopic nuclear matter properties,  in Fig.~\ref{Fig:param} we present the radii of neutron stars at $M=1.4 M_{\odot}$ and $2 M_{\odot}$ as well as the slope $L$ and curvature parameter $K_\mathrm{sym}$ of nuclear symmetry energy. It is evident that there are linear $L$-$K_\mathrm{sym}$ correlations in RMF models~\cite{Pais2012_PRL109-151101, Li2020_PRC102-045807}, where $K_\mathrm{sym}$ increases with $L$. For the macroscopic neutron star structures, it is found that $R_2$ generally coincides with $R_{1.4}$ except for the two cases obtained with the functionals TW99 and MTVTC, where the maximum masses are close to $2 M_{\odot}$ with $R_2<R_{1.4}$. In such cases, if the observed radii $R_{1.4}$ are indeed close to $R_2$ as observed in {NICER} and {XMM}-Newton missions~\cite{Riley2019_ApJ887-L21, Riley2021_ApJ918-L27, Miller2019_ApJ887-L24, Miller2021_ApJ918-L28}, then the maximum mass $M_\mathrm{max}$ of neutron stars could easily surpass $2.3 M_{\odot}$ as indicated in the lower-left panel of Fig.~\ref{Fig:param}, which approaches to the upper limit ($\le2.35 M_{\odot}$) according to the numerical simulations of binary neutron star merger event GW170817~\cite{Rezzolla2018_ApJ852-L25, Ruiz2018_PRD97-021501, Shibata2019_PRD100-023015}. Meanwhile, we note that the maximum mass $M_\mathrm{max}$ generally increases with radius, which reaches $2.77 M_{\odot}$ at $R_{1.4}= R_2=14.6$ km for the functional NL3. The linear correlations between neutron stars' radii and $L$ ($K_\mathrm{sym}$) are also observed. In the top-left panel of Fig.~\ref{Fig:param} we find $R_{1.4}$ increases with $L$, which is consistent with previous investigations that the radius and tidal deformability are closely related to $L$~\cite{Zhu2018_ApJ862-98, Tsang2019_PLB795-533, Dexheimer2019_JPG46-034002, Zhang2019_EPJA55-39, Zhang2020_PRC101-034303, Li2020_PRC102-045807}. At the same time, as indicated in the lower-right panel of Fig.~\ref{Fig:param}, the radii of two-solar-mass neutron stars $R_2$ seem to have a better correlation with the higher order coefficient $K_\mathrm{sym}$ instead of $L$, which is attributed to the larger density range covered in those stars. Such kind of correlations provide opportunities to constrain higher order coefficients of nuclear symmetry energy in the absence of strangeness via future radius measurements with both pulse-profile modeling~\cite{Riley2021_ApJ918-L27, Miller2021_ApJ918-L28} and gravitational wave observations~\cite{LVC2018_PRL121-161101}. Nevertheless, it is worth mentioning that the correlations are mainly due to the particular choices of covariant density functionals. If we consider a functional that does not follow the $L$-$K_\mathrm{sym}$ correlation, such as FSUGarnet~\cite{Parmar2022_PRD105-043017} indicated in Fig.~\ref{Fig:param}, the $K_\mathrm{sym}$-$R_{1.4,2}$ correlations become less evident than the $L$-$R_{1.4,2}$ correlations.

\section{\label{sec:con}Conclusion}
Based on the numerical recipe presented in our previous study~\cite{Xia2022_PRC105-045803}, in this work we investigate systematically the EOSs and microscopic structures of neutron star matter in a vast density range with $n_\mathrm{b}\approx 10^{-10}$-2 $\mathrm{fm}^{-3}$ adopting various covariant density functionals (NL3~\cite{Lalazissis1997_PRC55-540}, PK1~\cite{Long2004_PRC69-034319}, TM1~\cite{Sugahara1994_NPA579-557}, GM1~\cite{Glendenning1991_PRL67-2414}, and MTVTC~\cite{Maruyama2005_PRC72-015802}, DD-LZ1~\cite{Wei2020_CPC44-074107}, DDME-X~\cite{Taninah2020_PLB800-135065}, PKDD~\cite{Long2004_PRC69-034319}, DD-ME2~\cite{Lalazissis2005_PRC71-024312}, DD2~\cite{Typel2010_PRC81-015803}, and TW99~\cite{Typel1999_NPA656-331}). All the results are obtained in a unified manner adopting Thomas-Fermi approximation, where spherical and cylindrical symmetries are assumed for the WS cells. The optimum configurations of neutron star matter in $\beta$-equilibrium are obtained by searching for the energy minimum among six types of nuclear matter structures (droplet, rod, slab, tube, bubble, and uniform) at fixed baryon number density $n_\mathrm{b}$. The effects of charge screening are accounted for with electrons moving freely around the nucleus~\cite{Maruyama2005_PRC72-015802}, where the proton number of nucleus $Z$, droplet size $R_\mathrm{d}$, and WS cell size $R_\mathrm{W}$ become larger compared with the previous investigations neglecting the charge screening effects~\cite{Xia2022_PRC105-045803}. Note that we have adopted the SNA without any light clusters, which is not applicable for the functional DD2 as initially intended~\cite{Typel2010_PRC81-015803}. In such cases, we recommend Ref.~\cite{Fischer2014_EPJA50-46} for a more suitable EOS HS(DD2) obtained with the extended nuclear statistical equilibrium model.

The neutron drip densities of neutron star matter are found to be $n_\mathrm{d}\approx 2\text{-}3\times 10^{-4}$ fm${}^{-3}$, where those with the density-dependent couplings generally predict smaller $n_\mathrm{d}$ than that of non-linear ones. At smaller densities, neutron star matter are comprised of Coulomb lattices of nuclei and electrons with pressure mainly comes from electrons, where the EOSs of neutron star matter generally coincide with each other (discrepancy within 0.1\%). At $n_\mathrm{b}>n_\mathrm{d}$, the EOSs are sensitive to the adopted functionals, where the relative difference grows and reaches the peak at $n_\mathrm{b} \approx 0.02$ fm${}^{-3}$. The relative uncertainty of the EOSs decreases and remains small at $n_\mathrm{b} \lesssim 0.3$ fm${}^{-3}$, which however grows drastically at larger densities.

For the microscopic structures, it is found that only the droplet (crust) and uniform (core) phases emerge if the covariant density functionals with nonlinear self-couplings are adopted, while non-spherical shapes (rod, slab, tube, and bubble) may appear if density-dependent couplings are employed with generally smaller slope $L$ of symmetry energy. The corresponding core-crust transition densities $n_\mathrm{t}$ decreases with $L$ as well. Meanwhile, the obtained droplet size $R_\mathrm{d}$ and proton number of nucleus $Z$ approximately decrease with $L$, while the values of WS cell size $R_\mathrm{W}$ are close to each other. These observed trends generally coincide with previous investigations~\cite{Oyamatsu2007_PRC75-015801, Xu2009_ApJ697-1549, Grill2012_PRC85-055808, Bao2015_PRC91-015807, Shen2020_ApJ891-148, Xia2021_PRC103-055812}. Additionally, similar correlations with the curvature parameter $K_\mathrm{sym}$ are observed as well, which is closely related to the curvature-slope correlations~\cite{Pais2012_PRL109-151101, Li2020_PRC102-045807}.

The neutron star structures are then investigated adopting the unified EOSs. For all functionals considered in this work, the corresponding maximum masses of neutron stars exceed the two-solar-mass limit, while the functionals NL3, DD-LZ1, and DDME-X can even accommodate the mass of the secondary object observed in GW190814~\cite{Abbott2020_ApJ896-L44}. A combined constraint on both the masses and radii from pulsar observations~\cite{LVC2018_PRL121-161101, Riley2019_ApJ887-L21, Riley2021_ApJ918-L27, Miller2019_ApJ887-L24, Miller2021_ApJ918-L28} suggest that DD2, DD-LZ1, DD-ME2, and DDME-X are the most probable functionals for describing neutron star matter, while those of MTVTC and TW99 predict radii close to the lower ends of the PSR J0740+6620 constraints~\cite{Riley2021_ApJ918-L27, Miller2021_ApJ918-L28}. Nevertheless, in order to support massive neutron stars, the skewness coefficients $J$ for DD2, DD-LZ1, DD-ME2, and DDME-X are much larger than expected~\cite{Farine1997_NPA615-135, Xie2021_JPG48-025110}, which could be disentangled if the radius of PSR J0740+6620~\cite{Riley2021_ApJ918-L27, Miller2021_ApJ918-L28} and the maximum mass of neutron stars~\cite{Rezzolla2018_ApJ852-L25, Ruiz2018_PRD97-021501, Shibata2019_PRD100-023015} can be measured with higher accuracy. The functionals PKDD, GM1, TM1, PK1, and NL3 predict slightly too large radii according to the GW170817 constraint~\cite{LVC2018_PRL121-161101}, which can be reduced if exotic phases emerge at the center of neutron stars. Finally, we note there are approximate linear correlations between neutron stars' radii ($R_{1.4}$ at $M=1.4 M_{\odot}$ and $R_2$ at $M=2 M_{\odot}$) and the slope $L$ of nuclear symmetry energy. Since we have adopted covariant density functionals with approximate curvature-slope correlations, the correlations of those quantities with the curvature parameter $K_\mathrm{sym}$ of symmetry energy is observed as well.

It was shown that the neutron star structures are sensitive to the EOSs both in the core and crust regions, where a unified description for neutron star matter is required~\cite{Fortin2016_PRC94-035804, DinhThi2021_AA654-A114}. At the same time, the microscopic structures of neutron star matter play important roles in the corresponding transport and elastic properties, which affect various physical processes in neutron stars~\cite{Chamel2008_LRR11-10, Caplan2017_RMP89-041002}. Particularly, we have estimated the critical densities $n_\mathrm{DU}$ and neutron star masses $M_\mathrm{DU}$ at $Y_p=14.8\%$, above which the DU processes will take place and cool the neutron star too rapidly within just a few years~\cite{Klaehn2006_PRC74-035802, Page2006_NPA777-497}. We note that the DU processes only take place if functionals with $L\gtrsim 90$ MeV are adopted. The critical density lies in the range $n_\mathrm{DU}\approx0.23$-0.33 fm${}^{-3}>n_\mathrm{t}$, so that the DU processes is sensitive to the core EOSs. Meanwhile, the crust EOSs are closely connected to the fractional crustal moment of inertia ${I_\mathrm{c}}/{I}$, which can be constrained by the characteristic properties of glitches observed in pulsars. It is shown that ${I_\mathrm{c}}/{I}$ is sensitive to the adopted EOS and in particular the crust one, which provide opportunities to constrain neutron star structures and the corresponding EOS based on glitch monitoring. Further constraints may be obtained if we apply the current results to the investigations of other topics in pulsars such as asteroseismology~\cite{Kouveliotou1998_Nature393-235, Hurley1999_Nature397-41, Hansen1980_ApJ238-740, Schumaker1983_MNRAS203-457, McDermott1988_ApJ325-725,  Strohmayer1991_ApJ375-679, Passamonti2012_MNRAS419-638, Gabler2018_MNRAS476-4199, Sotani2012_PRL108-201101, Sotani2016_MNRAS464-3101, Kozhberov2020_MNRAS498-5149}, gravitational waves with respect to the strength of astromaterials~\cite{Horowitz2009_PRL102-191102, Chugunov2010_MNRAS407-L54, Horowitz2010_PRD81-103001, Caplan2018_PRL121-132701, Baiko2018_MNRAS480-5511, Abbott2020_ApJ902-L21}, neutrino-pasta scattering~\cite{Horowitz2016}, and evolution of magnetic field~\cite{Pons2013_NP9-431, Gao2017_ApJ849-19}. In such cases, the EOSs and microscopic structures of neutron star matter obtained in this work should be applicable for the investigations on the structures and evolutions of compact stars in a unified manner.

\section*{ACKNOWLEDGMENTS}
We would like to thank Prof. Nobutoshi Yasutake and Prof. Toshitaka Tatsumi for fruitful discussions. This work was supported by National SKA Program of China No.~2020SKA0120300, National Natural Science Foundation of China (Grant No.~11875052, No.~11873040, No.~11705163, and No.~11525524), the science research grants from the China Manned Space Project (No. CMS-CSST-2021-B11), the Youth Innovation Fund of Xiamen (No. 3502Z20206061), the Fundamental Research Funds for the Central Universities (Grant No.~lzujbky-2021-sp36), and the National Key R\&D Program of China No.~2018YFA0404402.


\newpage

%

\end{document}